# Universal Scaling Laws for Deep Indentation Beyond the Hertzian Regime

Tong Mu[1,2], Changhong Linghu[2,*], Yanju Liu[3], Jinsong Leng[1,†], Huajian Gao[4,§], K. Jimmy Hsia[2,5,¶]

[1]Centre of Composite Materials and Structures, Harbin Institute of Technology (HIT), Harbin 150080, China
[2]School of Mechanical and Aerospace Engineering, Nanyang Technological University, 50 Nanyang Avenue, Singapore 639798, Singapore
[3]Department of Astronautical Science and Mechanics, Harbin Institute of Technology (HIT), Harbin 150080, People's Republic of China
[4]Mechano-X Institute, Applied Mechanics Laboratory, Department of Engineering Mechanics, Tsinghua University, Beijing 100084, China
[5]School of Chemistry, Chemical Engineering and Biotechnology, Nanyang Technological University, 50 Nanyang Avenue, Singapore 639798, Singapore

Deep indentation of soft materials is ubiquitous across scales in nature and engineering, yet accurate predictions of contact behaviors under extreme deformations ($\delta/R > 1$) remain elusive due to geometric and material nonlinearities. Here, we investigate the indentation of rigid spheres into soft elastic substrates, resolving the highly nonlinear regime where the sphere becomes fully submerged. A universal geometric mapping approach reveals Hertz-type pressure distributions in the deformed configuration, validated by FEA. Closed-form solutions for contact force and radius agree with simulations up to $\delta/R = 2.5$. Experiments spanning soft polymers (Ecoflex, PDMS), food substrates (tofu), and biological tissues (octopus) validate the derived scaling law for hyperelastic materials. Our results establish a universal framework for extreme mechanical interactions, with applications in soft robotics, bioengineered systems, and tissue mechanics.

Indentation has been a fundamental issue in contact mechanics ever since the pioneering work by Hertz in 1882 [1]. Hertzian contact theory gained widespread recognition [2,3] by providing an elegant and general solution (details in Section S1). Its framework, based on infinitesimal deformation, has been extended to viscoelastic [4,5], adhesion [6,7], and sealing [8] problems in which the behaviors vary with time. However, its applicability remains restricted to small deformations, where the indentation depth $\delta$ is small compared to the indenter radius $R$ [9-14].

Recently, researchers have shown growing interest in the contact mechanics of soft materials, with applications in fields ranging from cell and tissue mechanics [15,16] to soft robotics [17], wearable sensors [18], and material parameter identification [9]. Unlike classical contact problems in metallic systems, in which indentation is typically shallow, deep indentation is both common and critical in biological materials, such as cells [16], skins [15,18,19], and other tissues [20,21], as well as in polymeric and responsive materials like hydrogels [22,23], shape memory polymers (SMPs) [24-26], and liquid crystalline elastomers (LCEs) [27]. In these materials, accurate modeling of deep indentation is essential for understanding contact behaviors and informing their design and analysis.

Modeling of the nonlinear, deep indentation is essential to understand the contact behaviors of soft materials under large deformation. Traditional approaches have primarily focused on moderate indentation depths (typically $\delta/R \leq 1$), where analytical models based on linear or weakly nonlinear elasticity are applicable. Sneddon [28] provided a foundational analytical solution for rigid hemisphere indenting on linear elastic substrates, capturing key differences between spherical and parabolic indenter systems. Sneddon's solution is accurate for shallow to moderate indentation depth ($\delta/R < 0.6$). To address material nonlinearity, Sabin et al. [29] and Giannakopoulos et al. [30] employed second-order elasticity for incompressible hyperelastic materials, deriving improved force–displacement relations. Du et al. [31] further refined these models by incorporating a quartic profile approximation, achieving better predictions of contact forces [29,30]. However, all these models are fundamentally restricted by their formulation in the undeformed configuration, preventing accurate description of the deformed contact geometry and limiting their applicability and possible extension to deep indentation ($\delta/R > 1$), where geometric nonlinearity becomes dominant.

Our experimental images in Fig. 1a show that the contact surface undergoes substantial deformation as the sphere gradually indents into the substrate. Although geometric nonlinearity has long been recognized in solid mechanics as essential for finite deformation, its role in contact mechanics, particularly for soft materials, remains underexplored. In this study, we investigate deep indentation behaviors through a combination of theory, simulations and experiments. We propose an analytical framework based on a geometric mapping, which relates the contact pressure distribution under large deformation to the classical Hertz theory. This approach provides a unified description of soft matter indentation for a wide range of materials and deformation scales, highlighting the critical role of geometric nonlinearity in contact mechanics of soft systems.

[*]Contact author: ling0090@e.ntu.edu.sg
[†]Contact author: lengjs@hit.edu.cn
[§]Contact author: gao.huajian@mail.tsinghua.edu.cn
[¶]Contact author: kjhsia@ntu.edu.sg

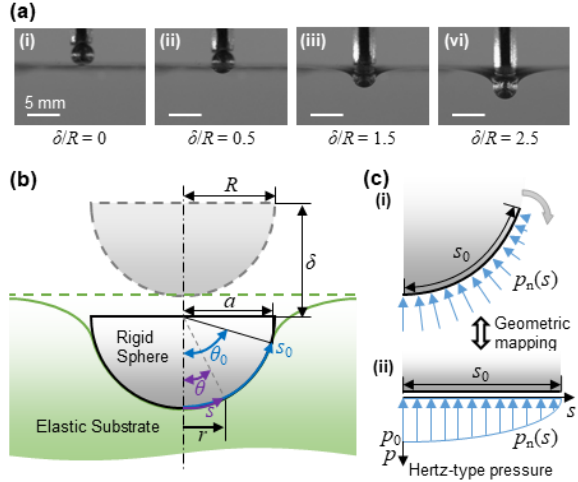

**FIG. 1. Deep indentation of a rigid sphere into a soft elastic substrate.** (**a**) Photographs of steel sphere indenting a PDMS substrate at various indentation depths: (i) $\delta/R = 0$, (ii) $\delta/R = 0.5$, (iii) $\delta/R = 1.5$, and (iv) $\delta/R = 2.5$. (**b**) Schematic configuration of a rigid sphere indenting an elastic substrate under finite deformation. (**c**) Geometric mapping of the normal pressure: (i) pressure distribution along the sphere arc, and (ii) mapping of the pressure distribution onto a straight line while preserving arc length, generating a Hertz-type pressure profile.

Different from the contact theory for small deformations, which assumes the difference between the deformed configuration and the undeformed configuration is neglectable, the two configurations under large deformations during deep indentation should be considered separately, as illustrated in Fig. 1b. In Hertz theory, the contact pressure is assumed to be perpendicular to the original contact plane and its distribution is described as a function of the contact radius, which equals to the arc length of the contact region under infinitesimal deformation. However, when the indentation depth is finite, the contact surface can no longer be assumed as planar, and the pressure is now perpendicular to the deformed surface. Therefore, the simplification regarding the pressure pointing vertically along the contact radius does not work anymore, as illustrated in Fig. 1c.

We now assume that the contact pressure $p_n$, i.e., the normal component of Cauchy stress on the contact surface in the deformed configuration, follows a Hertz-type distribution as a function of the arc length $s$, as illustrated in Fig.1c-I, as,

$$p_n(s) = p_0\left(1 - \frac{s^2}{s_0^2}\right)^{1/2}, (s \leq s_0), \quad (1)$$

where $p_0$ is the pressure at the center point of the sphere and $s_0$ is the arc length of the contact region. Defining central angles $\theta = s/R$ and $\theta_0 = s_0/R$, the pressure distribution is written as:

$$p_n(\theta) = p_0\left(1 - \frac{\theta^2}{\theta_0^2}\right)^{1/2}, (\theta \leq \theta_0), \quad (2)$$

To better visualize the pressure profile, we introduce a geometric mapping, as illustrated in Fig. 1c-ii, in which the contact arc is straightened while preserving its arc length, transforming the pressure distribution onto a flat surface. The pressure profile reduces exactly to the Hertz-type form derived under infinitesimal deformation. This correspondence establishes a direct analogy between deep indentation and the classical Hertz contact problem, and therefore we refer to this approach as the geometric mapping method.

For a semi-infinite body of linear elastic material with Young's modulus $E$ and Poisson's ratio $v$, the vertical displacement $u_z$ caused by a surface pressure distribution $p_z(x, y)$ in the vertical direction is given by the classic Boussinesq solution [32] and expressed as a convolution:

$$u_z(x,y) = -\frac{1}{\pi E^*}\int_{-\infty}^{+\infty}$$
$$\int_{-\infty}^{+\infty} \frac{p_z(x',y')}{\sqrt{(x-x')^2+(y-y')^2}}dx'dy'. \quad (3)$$

where the effective modulus $E^* = E / (1 - v^2)$. In the axisymmetric case, $r = \sqrt{x^2 + y^2}$. In the deep indentation problem, the vertical pressure of an infinitesimal ring becomes $p_z \cdot dr = p_n \cdot \cos\theta \cdot Rd\theta$. By using $r = R\sin\theta$ and $a = R\sin\theta_0$, $p_z$ is written as:

$$p_z(r) = p_0\sqrt{1 - \frac{\arcsin(r/R)^2}{\arcsin(a/R)^2}}, (r \leq a). \quad (4)$$

The vertical displacement caused by contact pressure given in Eq. (2) can be calculated according to Eq. (3). Although the integral does not have an analytical form, an approximate solution can be obtained as (details in Section S2):

$$\hat{u}_z = \frac{\sqrt{\pi}p_0}{4E^*A}(2A - A_r), \quad (5)$$

where $A$ is the total contact area, and $A_r$ is the contact area within radius $r$, as calculated by area formula of spherical crown as:

$$A_r = 2\pi R^2 (1 - \cos\theta), \quad (6)$$
$$A = 2\pi R^2 (1 - \cos\theta_0). \quad (7)$$

Furthermore, the vertical displacement within the contact region should satisfy the boundary condition:

$$u_z = \delta - R(1 - \cos\theta). \quad (8)$$

From Eqs. (5) and Eq. (8) and considering Eqs. (6) and (7), we have:

$$\frac{\sqrt{\pi}p_0}{4E^*A}[4\pi R^2(1-\cos\theta_0) - 2\pi R^2(1-\cos\theta)]$$
$$= \delta - R(1 - \cos\theta). \quad (9)$$

Since Eq. (9) is valid for any value of $\theta$, the terms in Eq. (9) consisting of $\cos\theta$ and those consisting only geometric and material constant must be independent, from which we obtain the maximum contact pressure and the contact angle as:

$$p_0 = \frac{2}{\pi}E^*\sqrt{\delta/R}, \quad (10)$$

$$\theta_0 = \arccos\left(1 - \frac{1}{2}\frac{\delta}{R}\right). \quad (11)$$

Additionally, because $a = R\sin\theta_0$, the contact radius is derived from Eq. (7) as:

$$a = \sqrt{\delta R - \frac{1}{4}\delta^2}. \quad (12)$$

To validate the stress distributions approximated in Eqs. (1) and (4) for deep indentation, finite element analysis (FEA) simulations (details in Section S3) were carried out using commercial software ABAQUS 2020 [33]. The substrate thickness is set to 30 times the indenter radius to eliminate finite-thickness effect (details in Section S4). A Neo-Hookean constitutive law, commonly used for hyperelastic materials, is adopted, with an initial shear modulus $\mu = 1$ MPa and Poisson's ratio $\nu = 0.5$.

Theoretical results are compared with the FEA results at indentation depths of $\delta/R = 0.5$, 1.0, and 2.0, as shown in Fig. 2. For comparison, the predictions from the classic Hertz theory are also included. Figure 2a shows that the Hertz theory predicts a self-similar pressure distribution along the radial direction, independent of indentation depth, which does not agree with the FEA results. In contrast, Figure 2 show that the proposed model accurately captures the normal pressure distributions (Eq. (1)) either in terms of radial coordinate $r/R$, or along the arc length $s/R$. Figure 2b demonstrate that pressure distribution along the arc length remains Hertz-like even at large deformations. The effective vertical pressure distributions obtained from the geometric mapping (Eq. (4)) agree well with FEA results (Fig. 2a), further validating the proposed framework for deep indentation.

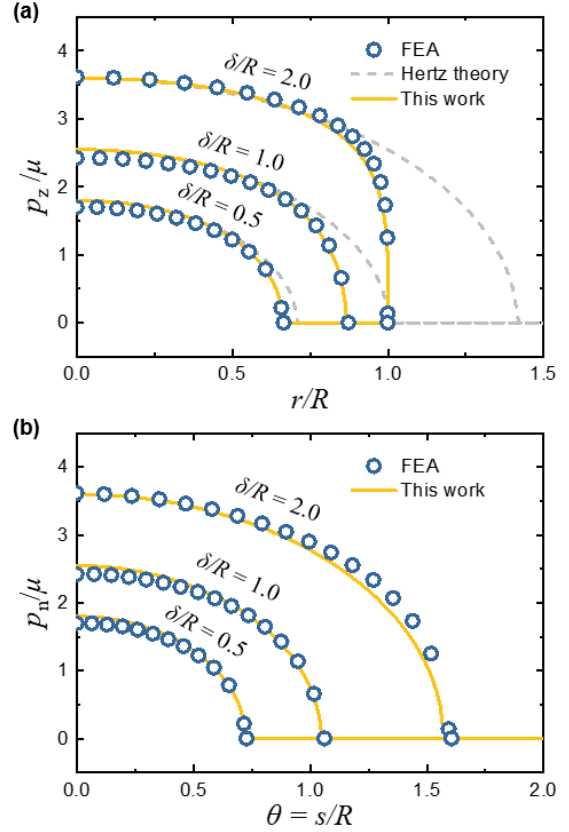

FIG. 2. **Contact pressure distributions in deep indentation.** (a) Normal pressure along the arc length. (b) Vertical component of pressure along the radial direction. Predictions from proposed model (yellow solid line, $E^* = 4$ MPa) are compared with FEA results (blue circles) at indentation depths of $\delta/R = 0.5$, 1.0, and 2.0. Hertz theory (gray dashed line) is included for reference. FEA simulations use a Neo-Hookean model with an initial shear modulus of 1 MPa and Poisson's ratio of 0.5.

The total contact force is obtained by integrating the vertical pressure components in Eq. (4) over the contact area, as:

$$F = \int_0^{\theta_0} p(\theta)\cos\theta\, 2\pi R\sin\theta\, R d\theta$$
$$= \frac{\pi^2}{4}p_0 R^2 H_1(2\theta_0), \quad (13)$$

where $H_1(x)$ is the Struve H function [34].

Substituting Eq. (10) into Eq. (13), the contact force is then expressed in terms of indentation depth as:

$$F = \frac{\pi}{2}E^*\sqrt{R^3\delta}\,H_1\left(2\arccos\left(1 - \frac{1}{2}\frac{\delta}{R}\right)\right) \quad (14)$$

Applying the approximation $H_1(x) \sim \frac{2}{3\pi}x^2$ near $x = 0$, Eq. (14) simplifies to:

$$F_H = \frac{4}{3}E^*\sqrt{R\delta^3} \quad (15)$$

where $F_H$ is Hertzian contact force.

Expanding Eq. (14) using Taylor series, the contact force is expressed as:

$$F = \frac{4}{3}E^*\sqrt{R\delta^3}\left[1 - \frac{11}{60}\frac{\delta}{R} - \frac{1}{350}\left(\frac{\delta}{R}\right)^2 + o\left(\left(\frac{\delta}{R}\right)^2\right)\right] \approx \left(1 - \frac{11}{60}\frac{\delta}{R}\right)F_H \quad (16)$$

The coefficient of the second-order term is negligibly small (details in Section S5), the contact force can be approximated by retaining only the first two terms as in Eq. (16).

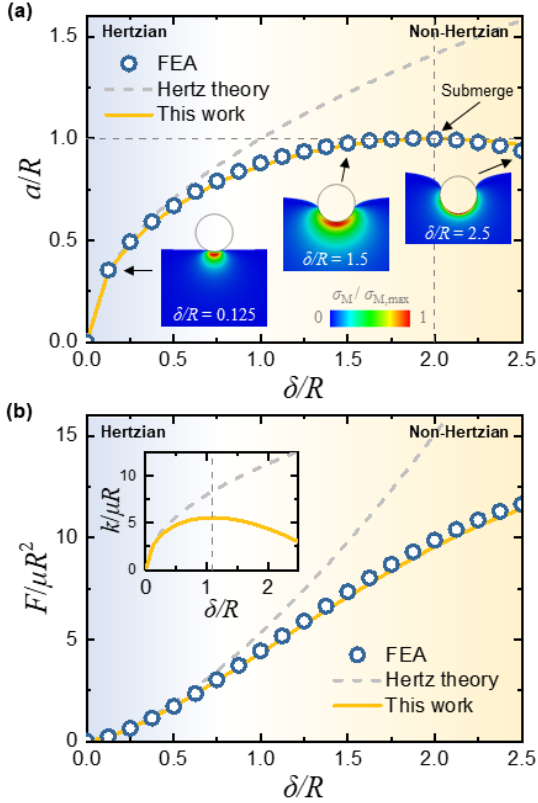

**FIG. 3. Predictions of the contact behaviors of deep indentation.** (**a**) Contact radius and (**b**) contact force versus indentation depth curves obtained from FEA results (blue circles) and predicted by the proposed model (yellow solid lines) and Hertz theory (gray dashed lines). Crosses indicate where the predicted contact radius exceeds the radius of the sphere in the models. The inset in (**b**) shows the contact stiffness as a function of the indentation depth.

Figure 3 compares the FEA results and theoretical predictions for indentation into a neo-Hookean solid substrate. The Hertz theory (gray dashed lines) quickly derivates from the FEA results (blue circles) once $\delta/R$ exceeds 0.125. The error between Hertz theory and FEA results exceeds 10% at $\delta/R = 1$ and reaches 60% at $\delta/R = 2.5$ for both contact force and contact radius. In contrast, our model (solid yellow lines) accurately captures the nonlinear behavior, aligning well with FEA results across the full range for both contact radius (Fig. 3a) and contact force (Fig. 3b). The maximum relative error remains below 3.0% for the contact radius and 6.5% for the contact force for indentation depth up to $\delta/R=2.5$. Extensive comparisons also highlight the better accuracy of our model compared with existing complex modified models [28-31,35,36] (details in Section S6).

Insets in Fig. 3a show the evolution of the contact configurations, visualized by the normalized von Mises stress distribution in the substrate at selected indentation depths. The stress distribution progressively loses self-similarity as the system transitions from Hertzian to non-Hertzian contact regimes, indicating the breakdown of Hertz theory under deep indentation. The vertical dashed line at $\delta/R = 2$ marks the point where the top of the sphere just submerges into the substrate. Notably, our model predicts that the lower hemisphere fully contacts the substrate at this point, after which the contact radius decreases.

Figure 3b gives the contact force as a function of indentation depth. Compared with the power function of Hertz theory, the curve exhibits an "S" shape, giving the contact stiffness as derivative of contact force as:

$$k = \frac{dF}{d\delta} = 2E^*\sqrt{R\delta}\left(1 - \frac{11}{36}\frac{\delta}{R}\right) \quad (17)$$

The inset in Fig. 3b shows the contact stiffness versus $\delta/R$, revealing a peak of $k = 0.9735\,E^*$ at $\delta/R = 11/12$ followed by a decline in stiffness.

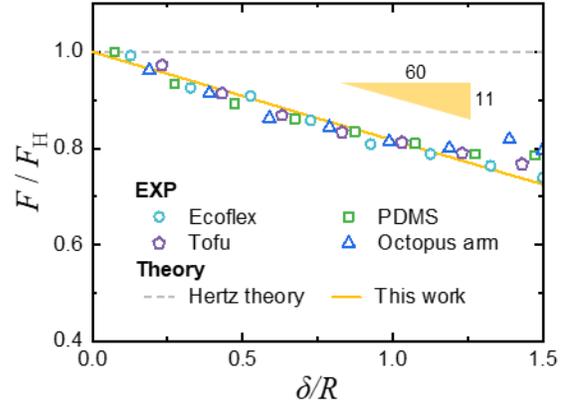

**FIG. 4.** Universal scaling law of the contact force under deep indentation for various materials, including commercial soft polymers (Ecoflex 00-30, and PDMS), food (Tofu), and biological tissue (Octopus tentacle).

In deep indentation, both geometric and material nonlinearity are potential sources of non-Hertz contact behavior. To evaluate the mechanical response of various soft materials under deep indentation, we present a comparison of experimentally measured results of $F/F_H$ versus the indentation depth for various materials (Fig. 4), including commercial soft polymers

(Ecoflex 00-30 and PDMS), food (tofu), and biological tissue (octopus tentacle). The results reveal a clear scaling behavior consistent with our model for all materials and a systematic deviation from the Hertzian predictions as indentation depth increases, indicating that the geometric nonlinearity effects are dominant in contact mechanics under large deformations. Despite vast differences in composition and structure, the results for all tested materials collapse onto a single curve given by Eq. (16), demonstrating a universal scaling law governing the deep indentation response for various soft materials (details in Section S7).

At large indentation depth ($\delta/R > 1.3$), the octopus arm deviates upward from the universal scaling due to strong material nonlinearity typical of biological tissues, such as J-curve stiffening [37]. PDMS also shows minor deviations, likely caused by tangential adhesion, which increases the contact force beyond theoretical predictions. These observations highlight that material non-linearity and adhesion warrant further investigation in the future.

In summary, we combine theory, experimentation, and simulations to resolve the deep indentation of rigid spheres into soft elastic substrates far beyond the classical Hertzian regime ($\delta/R > 1$). A geometric mapping approach uncovers a generalized Hertz-type pressure distribution along the contact arc, enabling accurate modeling of extreme deformations. Closed-form solutions for the contact pressure, force, and radius reveal that geometric nonlinearity—not material nonlinearity—dominates the response. The resulting universal scaling law, validated across polymers and biological tissues, provides a fundamental framework for soft material design, with immediate implications for soft robotics, biomedical devices, and tissue mechanics.

K.J.H., H.G., and C.L. acknowledge support from the Ministry of Education (MOE) of Singapore under the Academic Research Fund Tier 2 (MOE-T2EP50122-0001). Y.L. and J.L. acknowledges support from the National Key R&D Program of China (2022YFB3805700). T.M. acknowledges support from the China Scholarship Council program (202406120073).